\documentstyle[seceq,preprint]{ptptex}

\newcommand{\ket}[1]{| {#1} \rangle}     
\newcommand{\wtilde}[1]{\widetilde{#1}} 

\title{
Coupling Schemes for an $n$ $su(2)$ Spin System}

\author{
Masatoshi {\sc Yamamura},$^{1}$
Constan\c{c}a {\sc Provid\^encia},$^{2}$\\
Jo\~ao da {\sc Provid\^encia}$^{2}$, 
Yasuhiko {\sc Tsue}$^{3}$ 
and Jo\~ao da {\sc Provid\^encia}, Jr.$^{4}$, 

}

\inst{
$^1$Faculty of Engineering, Kansai University, Suita 564-8680, Japan\\
$^{2}$Departamento de Fisica, Universidade de Coimbra, P-3000 Coimbra, 
Portugal\\
$^{3}$Physics Division, Faculty of Science, Kochi University, Kochi 780-8520, 
Japan\\
$^{4}$Departamento de Fisica, Universidade de Beira Interior, 
6201-001 Covilh\~a, Portugal
}

\recdate{
\today
}

\abst{
In the framework of the Schwinger boson representation for the 
$su(2)$-algebra, the closed form is derived for the total spin 
eigenstates which result from the coupling of $n\ su(2)$-spins. 
In order to demonstrate its usefulness, the orthogonal set 
for the $so(5)$-algebra, which is reduced to four $su(2)$-spin 
systems, is obtained. 
}

\begin{document}

\maketitle

\section{Introduction and preliminaries}
The boson realization of the $su(2)$-algebra due to Schwinger\cite{S55} 
and named
after him plays a crucial role in the Theory of Angular Momentum,
for instance in the derivation of the irreducible representations
of the rotation operators. The Schwinger representation of the
$su(2)$-algebra also proved to be most useful in the study of
various many-body systems, for example, shown in Ref. \citen{KPTYsup}. 
Appropriate extensions of the Schwinger
representation have been obtained by two of the present authors
(J.P. and M.Y.) and Kuriyama and applied to solve diverse problems
formulated in the framework of the $su(M+1)$ and 
the $su(N,1)$-algebra.\cite{KPY00}
In the present paper, the Schwinger representation is applied to
derive closed form expressions describing the total spin
eigenstates which result from the coupling of $n$ $su(2)$-spins.
The obtained expressions may be of interest in the theoretical
investigation of spin systems, for instance, the description of the Heisenberg 
antiferromagnet, the simple case of which was recently given by 
four of the present authors (J.P., M.Y., C.P. and J.P.Jr.) and 
Brajczewska.\cite{PY04} 
But, it may be also useful for the description of many-body systems 
which can be reduced to many $su(2)$-spins, 
for example, the $so(5)$-algebra which plays an important role in the 
studies of the charge-independent pairing correlations in nuclei\cite{a5} 
and the high-temperature superconductivity.\cite{a6}

It is well known that the $su(2)$-algebra is composed of three operators, 
${\hat S}_{\pm,0}$, which are called the $su(2)$-generators and obey 
the following relations : 
\begin{eqnarray}
& &{\hat S}_+^*={\hat S}_- \ , \qquad {\hat S}_0^*={\hat S}_0 \ , 
\label{1}\\
& &[\ {\hat S}_+ , {\hat S}_-\ ]=2{\hat S}_0\ , \qquad
[\ {\hat S}_0 , {\hat S}_{\pm}\ ]=\pm{\hat S}_{\pm} \ . 
\label{2}
\end{eqnarray}
Hereafter, we express the set $({\hat S}_{\pm,0})$ also ${\hat {\mib S}}$, 
and then the Casimir operator can be expressed in the form 
\begin{subequations}
\begin{equation}\label{3}
{\hat {\mib S}}^2={\hat S}_0^2+({\hat S}_-{\hat S}_+ 
+ {\hat S}_+{\hat S}_-)/2 \ .
\end{equation}
\end{subequations}
The eigenvalue problem of the $su(2)$-algebra is formulated in terms of 
the relations 
\begin{subequations}\label{4}
\begin{eqnarray}
& &{\hat {\mib S}}^2 \ket{(\alpha);ss_0}=s(s+1)\ket{(\alpha);ss_0} \ , 
\quad (s=0,1/2,1,3/2,\cdots) 
\label{4a}\\
& &{\hat {S}_0} \ket{(\alpha);ss_0}=s_0\ket{(\alpha);ss_0} \ , 
\quad (s_0=-s, -s+1, \cdots s-1, s) 
\label{4b}
\end{eqnarray}
\end{subequations}
The symbol $(\alpha)$ denotes a set of the quantum numbers additional to 
$(s, s_0)$. If the operator ${\hat S}$, which gives us 
${\hat {\mib S}}^2={\hat S}({\hat S}+1)$, can be defined, the relation 
(\ref{4a}) is reduced to ${\hat S}\ket{(\alpha);ss_0}=s\ket{(\alpha);ss_0}$. 
In this paper, we call the system obeying the above algebra 
$su(2)$-spin system. 
In this sense, ${\hat S}$ plays a role of the magnitude of the $su(2)$-spin.

The simplest boson realization of the $su(2)$-algebra was presented by 
Schwinger\cite{S55} and it is called the Schwinger boson representation. 
This realization is formulated in terms of two kinds of bosons 
$({\hat a},{\hat a}^*)$ and $({\hat b},{\hat b}^*)$. 
The set ${\hat {\mib S}}$ is given as 
\begin{equation}\label{5}
{\hat S}_+={\hat a}^*{\hat b}\ , \quad 
{\hat S}_-={\hat b}^*{\hat a}\ , \quad 
{\hat S}_0=({\hat a}^*{\hat a}-{\hat b}^*{\hat b})/2\ .
\end{equation}
The operator ${\hat {\mib S}}^2$ can be expressed as 
\begin{equation}\label{6}
{\hat {\mib S}}^2={\hat S}({\hat S}+1) \ , \qquad
{\hat S}=({\hat a}^*{\hat a}+{\hat b}^*{\hat b})/2\ .
\end{equation}
The eigenstate $\ket{ss_0}$ (in this case, $(\alpha)$ is not necessary) 
can be expressed as 
\begin{equation}\label{7}
\ket{ss_0}=({\hat S}_+)^{s+s_0}({\hat b}^*)^{2s}\ket{0} \ . 
\end{equation}
Here, the normalization constant is omitted, and hereafter, we 
will omit it for any state. 
The state $\ket{ss_0}$ satisfies 
\begin{equation}\label{8}
{\hat S}\ket{ss_0}=s\ket{ss_0} \ . 
\end{equation}
Clearly, with the aid of the above treatment, we can describe single 
$su(2)$-spin system. 
It should be noted that the operator ${\hat S}$, which plays a role of the 
magnitude of the $su(2)$-spin, can be defined only in the case of 
single $su(2)$-spin. 
The above is our starting arguments. 

In the next section, the cases of two and three $su(2)$-spins are 
recapitulated. 
Sections 3 and 4 are the high light parts of this paper, where following 
two types of the coupling orders for the $su(2)$-spins, the eigenstates 
of total spin are obtained. 
In \S 5, the coupling rules for the two orders are discussed. 
Section 6 is devoted to the application of the coupling scheme presented in 
this paper to the $so(5)$-algebra, which is reduced to four $su(2)$-spin 
systems. Finally, in \S 7, three comments are mentioned. 
Especially, in relation to the charge-independent pairing theory, 
the eigenstates given in \S 6 is reexpressed.

\section{The cases of two and three $su(2)$-spins}

A possible Schwinger boson representation of two and three $su(2)$-spin 
systems was formulated by the present authors and Kuriyama 
in Refs. \citen{22} and 
\citen{3}, respectively. 
The former and the latter are referred to as (A) and (B), respectively. 
It may be helpful for readers to recapitulate the 
formalism in terms of the notations used in the original papers.

The case (A) is formulated in terms of the bosons $({\hat a}_+,{\hat b}_+, 
{\hat a}_-, {\hat b}_-)$ and the following $su(2)$-generators plays a 
central role : 
\begin{eqnarray}
& &{\hat S}_{\pm,0}={\hat I}_{\pm,0}+{\hat J}_{\pm,0} \ , \quad
{\rm i.e.,}\quad 
{\hat {\mib S}}={\hat {\mib I}}+{\hat {\mib J}} \ , 
\label{9}\\
& &{\hat I}_+={\hat a}_+^*{\hat b}_+ \ , \qquad
{\hat I}_-={\hat b}_+^*{\hat a}_+ \ , \qquad
{\hat I}_0=({\hat a}_+^*{\hat a}_+-{\hat b}_+^*{\hat b}_+)/2 \ , \nonumber\\
& &{\hat J}_+={\hat a}_-^*{\hat b}_- \ , \qquad
{\hat J}_-={\hat b}_-^*{\hat a}_- \ , \qquad
{\hat J}_0=({\hat a}_-^*{\hat a}_--{\hat b}_-^*{\hat b}_-)/2 \ . 
\label{10}
\end{eqnarray}
The sets ${\hat {\mib I}}$ and ${\hat {\mib J}}$ also form the 
$su(2)$-algebras, respectively, and the Casimir operators are expressed as 
\begin{eqnarray}\label{11}
& &{\hat {\mib I}}^2={\hat I}({\hat I}+1) \ , \qquad
{\hat I}=({\hat a}_+^*{\hat a}_++{\hat b}_+^*{\hat b}_+)/2 \ , \nonumber\\
& &{\hat {\mib J}}^2={\hat J}({\hat J}+1) \ , \qquad
{\hat J}=({\hat a}_-^*{\hat a}_-+{\hat b}_-^*{\hat b}_-)/2 \ .
\end{eqnarray}
In (A), the operators ${\hat T}_+$ and ${\hat R}_+$ defined in the following 
also play an important role : \break
\begin{equation}\label{12}
{\hat T}_+={\hat a}_+^*{\hat b}_-^*-{\hat b}_+^*{\hat a}_- \ , \qquad
{\hat R}_+={\hat a}_+^*{\hat a}_-+{\hat b}_+^*{\hat b}_- \ . 
\end{equation}
In the present system, mutually commutable operators are ${\hat I}$, 
${\hat J}$, ${\hat {\mib S}}^2$ and ${\hat S}_0$, and then, we can set up 
the eigenvalue equation 
\begin{eqnarray}\label{13}
& &{\hat I}\ket{ij;ss_0}=i\ket{ij;ss_0} \ , \qquad
{\hat J}\ket{ij;ss_0}=j\ket{ij;ss_0} \ , \nonumber\\
& &{\hat {\mib S}}^2\ket{ij;ss_0}=s(s+1)\ket{ij;ss_0} \ , \qquad
{\hat S}_0\ket{ij;ss_0}=s_0\ket{ij;ss_0} \ . 
\end{eqnarray}
The eigenstate $\ket{ij;ss_0}$ is given in the form 
\begin{equation}\label{14}
\ket{ij;ss_0}=({\hat S}_+)^{s+s_0}({\hat T}_+)^{i+j-s}({\hat R}_+)^{i-j+s} 
({\hat b}_-^*)^{2s}\ket{0} \ . 
\end{equation}
The set of the quantum numbers $ij$ is just $(\alpha)$. 
The form (\ref{14}) is found in Eq.(A$\cdot$2$\cdot$27) in a form 
slightly different ordering of the operators from that of the present. 
For the state (\ref{14}), the quantum numbers obey 
\begin{subequations}\label{15}
\begin{eqnarray}
& &2s \geq 0 \ , \quad s+s_0 \geq 0 \ , \quad 2s-(s+s_0) \geq 0 \ , 
\label{15a}\\
& &i+j-s \geq 0 \ , \quad i-j+s \geq 0 \ , \quad 2s-(i-j+s) \geq 0 \ .
\label{15b}
\end{eqnarray}
\end{subequations}
The relation (\ref{15}) is derived under the condition that the 
state (\ref{14}) should not vanish. 
From the relation (\ref{15}), we obtain 
\begin{equation}\label{16}
s \geq 0 \ , \qquad |s_0| \leq s \ , \qquad 
|i-j| \leq s \leq i+j \ . 
\end{equation}
The above is identical with the coupling rule of two $su(2)$-spins.

The case (B) is formulated in terms of bosons 
$({\hat a}_i , {\hat b}_i ; i=1,2,3)$ and ${\hat S}_{\pm,0}$ can be 
expressed as 
\begin{eqnarray}
& &{\hat S}_{\pm,0}=\sum_{i=1}^{3} {\hat S}_{\pm,0}(i) \ , \quad 
{\rm i.e.,}\quad {\hat {\mib S}}=\sum_{i=1}^{3}{\hat {\mib S}}(i) \ , 
\label{17}\\
& &{\hat S}_+(i)={\hat a}_i^*{\hat b}_i \ , \qquad
{\hat S}_-(i)={\hat b}_i^*{\hat a}_i \ , \qquad
{\hat S}_0(i)=({\hat a}_i^*{\hat a}_i-{\hat b}_i^*{\hat b}_i)/2 \ . 
\label{18}
\end{eqnarray}
Each of the sets $({\hat {\mib S}}(i) ; i=1,2,3)$ also forms the 
$su(2)$-algebra and the Casimir operator can be expressed as 
\begin{equation}\label{19}
{\hat {\mib S}}(i)^2={\hat S}(i)({\hat S}(i)+1) \ , \qquad
{\hat S}(i)=({\hat a}_i^*{\hat a}_i+{\hat b}_i^*{\hat b}_i)/2 \ .
\end{equation}
In (B), the following operators also play an important role : 
\begin{eqnarray}
& &{\hat Q}_+={\hat a}_1^*{\hat b}_2^*-{\hat b}_1^*{\hat a}_2^* \ , 
\qquad
{\hat M}_+={\hat a}_1^*{\hat a}_2+{\hat b}_1^*{\hat b}_2 \ , 
\nonumber\\
& &{\hat T}_+(2)={\hat a}_2^*{\hat b}_3^*-{\hat b}_2^*{\hat a}_3^* \ , 
\qquad
{\hat R}_+(2)={\hat a}_2^*{\hat a}_3+{\hat b}_2^*{\hat b}_3 \ . 
\label{20}
\end{eqnarray}
In the present system, mutually commutable operators are 
${\hat S}(i)$ ($i=1,2,3$), ${\hat {\mib S}}_{12}^2$, ${\hat {\mib S}}^2$ 
and ${\hat S}_0$. 
Here, ${\hat {\mib S}}_{12}^2$ denotes 
\begin{eqnarray}\label{21}
{\hat {\mib S}}_{12}^2=({\hat S}_0(1)+{\hat S}_0(2))^2
&+&
\frac{1}{2}
\biggl[ ({\hat S}_-(1)+{\hat S}_-(2))({\hat S}_+(1)+{\hat S}_+(2))
\nonumber\\
& &\ \ \ +({\hat S}_+(1)+{\hat S}_+(2))({\hat S}_-(1)+{\hat S}_-(2))
\biggl] \ . 
\end{eqnarray}
We can set up the following eigenvalue equation : 
\begin{eqnarray}\label{22}
& &{\hat S}(i)\ket{s_1s_2(s_{12})s_3;ss_0}
=s_i\ket{s_1s_2(s_{12})s_3;ss_0} \ , \quad (i=1,2,3) \nonumber\\
& &{\hat {\mib S}}_{12}^2\ket{s_1s_2(s_{12})s_3;ss_0}
=s_{12}(s_{12}+1)\ket{s_1s_2(s_{12})s_3;ss_0} \ , \nonumber\\
& &{\hat {\mib S}}^2\ket{s_1s_2(s_{12})s_3;ss_0}
=s(s+1)\ket{s_1s_2(s_{12})s_3;ss_0} \ , \nonumber\\
& &{\hat {S}}_{0}\ket{s_1s_2(s_{12})s_3;ss_0}
=s_{0}\ket{s_1s_2(s_{12})s_3;ss_0} \ . 
\end{eqnarray}
The eigenstate $\ket{s_1s_2(s_{12})s_3;ss_0}$ is obtained in 
the form 
\begin{eqnarray}\label{23}
\ket{s_1s_2(s_{12})s_3;ss_0}
&=&({\hat S}_+)^{s+s_0}({\hat Q}_+)^{s_1+s_2-s_{12}}
({\hat M}_+)^{s_1-s_2+s_{12}}\nonumber\\
& &\times ({\hat T}_+(2))^{s_{12}+s_3-s}({\hat R}_+(2))^{s_{12}-s_3+s}
({\hat b}_3^*)^{2s}\ket{0} \ . 
\end{eqnarray}
The form (\ref{23}) is found in Eq.(B$\cdot$5$\cdot$8) in a form
slightly different ordering of the operators from the present one. 
In the present case, $(\alpha)$ is just $(s_1s_2(s_{12})s_3)$. 
For the state (\ref{23}), the quantum numbers obey 
\begin{subequations}\label{24}
\begin{eqnarray}
& &2s \geq 0 \ , \qquad s+s_0 \geq 0 \ , \qquad 2s-(s+s_0) \geq 0 \ , 
\label{24a}\\
& &s_1+s_2-s_{12} \geq 0 \ , \qquad s_1-s_2+s_{12} \geq 0 \ , \nonumber\\
& &-(s_1-s_2+s_{12})+(s_{12}+s_3-s)+(s_{12}-s_3+s) \geq 0 \ , 
\label{24b}\\
& &s_{12}+s_3-s \geq 0 \ , \qquad s_{12}-s_3+s \geq 0 \ , 
\nonumber\\
& &2s-(s_{12}-s_3+s) \geq 0 \ . 
\label{24c}
\end{eqnarray}
\end{subequations}
The relation (\ref{24}) is derived under the condition that the state 
(\ref{23}) should not vanish. The relation (\ref{24}) gives us 
the coupling rule of three $su(2)$-spins : 
\begin{eqnarray}\label{25}
& &s \geq 0 \ , \qquad |s_0| \leq s\ , \nonumber\\
& &|s_1-s_2| \leq s_{12} \leq s_1+s_2\ , \qquad
|s_{12}-s_3| \leq s \leq s_{12}+s_3 \ .
\end{eqnarray}

\section{General case. I}

The aim of this paper is to provide a possible coupling scheme for 
$n$ $su(2)$-spin system by generalizing the cases of two and three 
$su(2)$-spin systems recapitulated in \S 2. 
The system under investigation consists of the $su(2)$-spins 
${\hat {\mib S}}(i)$ ($i=1,2,\cdots ,n)$. Then, the set ${\hat {\mib S}}$ 
is given as 
\begin{equation}\label{26}
{\hat {\mib S}}=\sum_{i=1}^{n}{\hat {\mib S}}(i) \ . 
\end{equation}
In this case, it may be indispensable to specify the order of 
the coupling. 
We discuss two cases. Our first task is to discuss the 
case of the following coupling order : 
\begin{equation}\label{27}
\sum_{i=1}^{n}{\hat {\mib S}}(i)=\left[\cdots\left[
[{\hat {\mib S}}(1)+{\hat {\mib S}}(2)]
+{\hat {\mib S}}(3)\right]+\cdots +
{\hat {\mib S}}(n-1)\right]+{\hat {\mib S}}(n)\ . 
\end{equation}
For example, in the case $n=4$, $\sum_{i=1}^{4}{\hat {\mib S}}(i)=
[[{\hat {\mib S}}(1)+{\hat {\mib S}}(2)]+{\hat {\mib S}}(3)]
+{\hat {\mib S}}(4)$. 
For this purpose, we introduce $2n$ kinds of boson operators : 
$({\hat a}_i , {\hat a}_i^*)$ and $({\hat b}_i , {\hat b}_i^*)$ 
$(i=1,2,\cdots , n)$. 
The $i$-th $su(2)$-spin ${\hat {\mib S}}(i)$ can be given as 
\begin{eqnarray}
& &{\hat S}_+(i)={\hat a}_i^*{\hat b}_i \ , \qquad
{\hat S}_-(i)={\hat b}_i^*{\hat a}_i \ , \qquad
{\hat S}_0(i)=({\hat a}_i^*{\hat a}_i-{\hat b}_i^*{\hat b}_i)/2 \ , 
\label{28}\\
& &{\hat S}(i)=({\hat a}_i^*{\hat a}_i+{\hat b}_i^*{\hat b}_i)/2\ . 
\label{29}
\end{eqnarray}
Further, we define the following operators : 
\begin{eqnarray}\label{30}
& &{\hat {\mib S}}[l]=\sum_{i=1}^{l}{\hat {\mib S}}(i) \ , \quad
{\rm i.e.,}\quad 
{\hat S}_{\pm,0}[l]=\sum_{i=1}^{l}{\hat S}_{\pm,0}(i) \ , \ \ (l=1,2,\cdots ,n)
\nonumber\\
& &{\hat {\mib S}}[l]^2={\hat S}_0[l]^2+({\hat S}_-[l]{\hat S}_+[l]
+{\hat S}_+[l]{\hat S}_-[l])/2 \ , 
\end{eqnarray}
\vspace{-0.8cm}
\begin{subequations}\label{31}
\begin{eqnarray}
& &{\hat X}_+(l)={\hat a}_l^*{\hat b}_{l+1}^*-{\hat b}_l^*{\hat a}_{l+1}^* \ , 
\label{31a}\\
& &{\hat Y}_+(l)={\hat a}_l^*{\hat a}_{l+1}+{\hat b}_l^*{\hat b}_{l+1} \ , 
\ \ (l=1,2,\cdots ,n-1) \qquad\qquad\qquad
\label{31b}
\end{eqnarray}
\end{subequations}
Clearly, ${\hat {\mib S}}[l]$ for $l=1,n$ are 
\begin{equation}\label{32}
{\hat {\mib S}}[1]={\hat {\mib S}}(1) \ , \qquad
{\hat {\mib S}}[n]={\hat {\mib S}} \ . 
\end{equation}
We change the original notations used in the recapitulation 
for $n=2$ and 3 : \\
\noindent
For (A), 
\begin{subequations}\label{33}
\begin{eqnarray}
& &{\hat a}_+ \rightarrow {\hat a}_1 \ , \quad 
{\hat b}_+ \rightarrow {\hat b}_1 \ , \quad 
{\hat a}_- \rightarrow {\hat a}_2 \ , \quad 
{\hat b}_- \rightarrow {\hat b}_2 \ , \nonumber\\
& &{\hat I}_{\pm,0} \rightarrow {\hat S}_{\pm,0}(1) \ , \quad
{\hat I} \rightarrow {\hat S}(1) \ , \quad
{\hat J}_{\pm,0} \rightarrow {\hat S}_{\pm,0}(2) \ , \quad
{\hat J} \rightarrow {\hat S}(2) \ , \nonumber\\
& &{\hat T}_+ \rightarrow {\hat X}_+(1)
\ , \qquad
{\hat R}_+ \rightarrow {\hat Y}_+(1)
\ , \nonumber\\
& &i+j-s \rightarrow 2\lambda_1 \ , \qquad 
i-j+s \rightarrow 2\mu_1 \ , 
\label{33a}\\
& &\ket{ij;ss_0} \rightarrow \ket{\lambda_1 \mu_1;ss_0}
=({\hat S}_+)^{s+s_0}({\hat X}_+(1))^{2\lambda_1}({\hat Y}_+(1))^{2\mu_1}
({\hat b}_2^*)^{2s}\ket{0} \ .
\label{33b}
\end{eqnarray}
\end{subequations}
For (B),
\begin{subequations}\label{34} 
\begin{eqnarray}
& &{\hat Q}_+ \rightarrow {\hat X}_+(1)
\ , \quad
{\hat M}_+ \rightarrow {\hat Y}_+(1)
\ , \quad
{\hat T}_+(2) \rightarrow {\hat X}_+(2)
\ , \quad
{\hat R}_+(2) \rightarrow {\hat Y}_+(2)
\ , \nonumber\\
& &{\hat S}_{\pm,0}(1)+{\hat S}_{\pm,0}(2) \rightarrow 
{\hat S}_{\pm,0}[2] \ , \quad
{\hat {\mib S}}_{12}^2 \rightarrow {\hat {\mib S}}[2]^2 \ , \nonumber\\
& &s_1+s_2-s_{12} \rightarrow 2\lambda_1 \ , \qquad 
s_1-s_2+s_{12} \rightarrow 2\mu_1 \ , \nonumber\\
& &s_{12}+s_3-s \rightarrow 2\lambda_2 \ , \qquad 
s_{12}-s_3+s \rightarrow 2\mu_2 \ , 
\label{34a}\\
& &\ket{s_1s_2(s_{12})s_3;ss_0} \rightarrow 
\ket{\lambda_1 \mu_1 \lambda_2 \mu_2;ss_0}\nonumber\\
& &\qquad\qquad\qquad\qquad
=({\hat S}_+)^{s+s_0}({\hat X}_+(1))^{2\lambda_1}({\hat Y}_+(1))^{2\mu_1}
({\hat X}_+(2))^{2\lambda_2}({\hat Y}_+(2))^{2\mu_2}
({\hat b}_3^*)^{2s}\ket{0} \ . \nonumber\\
& &
\label{34b}
\end{eqnarray}
\end{subequations}
The comparison between the two changes (\ref{33}) and (\ref{34}) 
gives us an idea for the generalization from the cases of two 
and three $su(2)$-spins. 
As a natural generalization of the states $\ket{\lambda_1 \mu_1 ; ss_0}$ 
and $\ket{\lambda_1 \mu_1 \lambda_2 \mu_2;ss_0}$ defined in 
the relations (\ref{33b}) 
and (\ref{34b}), respectively, it may be possible to define the state 
\begin{eqnarray}\label{35}
\ket{\lambda_1 \mu_1 \cdots \lambda_{n\!-\!1} \mu_{n\!-\!1};ss_0}&=&
({\hat S}_+)^{s+s_0}({\hat X}_+(1))^{2\lambda_1}({\hat Y}_+(1))^{2\mu_1}
\nonumber\\
&\times&\! 
\cdots\!\times\! ({\hat X}_+(n\!-\!1))^{2\lambda_{n\!-\!1}}
({\hat Y}_+(n-1))^{2\mu_{n-1}}
({\hat b}_n^*)^{2s} \ket{0} \ . \qquad\quad
\end{eqnarray}

In the present system, there exist $2n$ mutually commutable operators 
${\hat S}(l)$ $(l=1,2,\cdots, n)$, ${\hat {\mib S}}[l]^2$ 
$(l=2,3,\cdots, n-1)$, ${\hat {\mib S}}^2$ and ${\hat S}_0$. 
Our task is to prove that the state 
(\ref{35}) is the eigenstate for these operators. 
First, we show the case of ${\hat S}(l)$. 
For this task, the following relations are useful : 
\begin{subequations}\label{36}
\begin{eqnarray}
& &[\ {\hat S}(l)\ , \ {\hat X}_+(l-1) \ ]=(1/2){\hat X}_+(l-1) \ , 
\quad
[\ {\hat S}(l)\ , \ {\hat X}_+(l) \ ]=(1/2){\hat X}_+(l) \ , 
\nonumber\\
& &[\ {\hat S}(l)\ , \ {\hat Y}_+(l-1) \ ]=-(1/2){\hat Y}_+(l-1) \ , 
\quad
[\ {\hat S}(l)\ , \ {\hat Y}_+(l) \ ]=(1/2){\hat Y}_+(l) \ , 
\nonumber\\
& &[\ {\rm the\ other\ commbinations} \ ]=0 \ , 
\label{36a}\\
& &{\hat S}(n)({\hat b}_n^*)^{2s}\ket{0}=s({\hat b}_n^*)^{2s}\ket{0} \ . 
\label{36b}
\end{eqnarray}
\end{subequations}
With the use of the relation (\ref{36}), we can complete 
the above task and the 
eigenvalue of ${\hat S}(l)$, which we denote as $s_l$, is obtained as 
\begin{eqnarray}\label{37}
& &s_1=\lambda_1+\mu_1 \ , \nonumber\\
& &s_l=(\lambda_{l-1}-\mu_{l-1})+(\lambda_l+\mu_l)\ , 
\quad (l=2,3,\cdots,n-1) \nonumber\\
& &s_n=(\lambda_{n-1}-\mu_{n-1})+s \ . 
\end{eqnarray}
Next, we prove that the state (\ref{35}) is the eigenstate of 
${\hat {\mib S}}[l]^2$. 
For this purpose, first, the following relations should be noted : 
\begin{subequations}\label{38}
\begin{eqnarray}
& &[\ {\hat S}_+[l]\ , \ {\hat X}_+(l) \ ]
=-{\hat a}_l^*{\hat a}_{l+1}^* \ , \qquad
[\ {\hat S}_-[l]\ , \ {\hat X}_+(l) \ ]
={\hat b}_l^*{\hat b}_{l+1}^* \ , \nonumber\\
& &[\ {\hat S}_0[l]\ , \ {\hat X}_+(l) \ ]
=(1/2)({\hat a}_l^*{\hat b}_{l+1}^* - {\hat b}_l^*{\hat a}_{l+1}^*) \ , 
\nonumber\\
& &[\ {\hat S}_+[l]\ , \ {\hat Y}_+(l) \ ]
={\hat a}_l^*{\hat b}_{l+1} \ , \qquad
[\ {\hat S}_-[l]\ , \ {\hat Y}_+(l) \ ]
={\hat b}_l^*{\hat a}_{l+1} \ , \nonumber\\
& &[\ {\hat S}_0[l]\ , \ {\hat Y}_+(l) \ ]
=(1/2)({\hat a}_l^*{\hat a}_{l+1} - {\hat b}_l^*{\hat b}_{l+1}) \ , 
\nonumber\\
& &[\ {\rm the\ other\ combinations}\ ]=0 \ , 
\label{38a}\\
& &{\hat S}_{\pm,0}[l]({\hat b}_n^*)^{2s} \ket{0}=0 \ . 
\qquad
(l=2,3,\cdots, n-1) 
\label{38b}
\end{eqnarray}
\end{subequations}
With the use of the relation (\ref{38a}) with the definition of 
${\hat {\mib S}}[l]^2$, we obtain 
\begin{eqnarray}\label{39}
& &[\ {\hat {\mib S}}[l]^2\ , 
\ ({\hat X}_+(l))^{2\lambda_l}({\hat Y}_+(l))^{2\mu_l} \ ] \nonumber\\
& &\ \ = (\lambda_l+\mu_l)[(\lambda_l+\mu_l)+1]({\hat X}_+(l))^{2\lambda_l}
({\hat Y}_+(l))^{2\mu_l} +2{\hat Z}_+(l;\lambda_l\mu_l)\ , \nonumber\\
& &{\hat Z}_+(l;\lambda_l\mu_l) \nonumber\\
& &\ \ =
\biggl[ \lambda_l ({\hat X}_+(l))^{2\lambda_{l}-1}({\hat Y}_+(l))^{2\mu_l}
({\hat a}_l^*{\hat b}_{l+1}^*+{\hat b}_l^*{\hat a}_{l+1}^*) 
\nonumber\\
& &\ \ \ \ \ \ 
+\mu_l({\hat a}_l^*{\hat a}_{l+1}-{\hat b}_l^*{\hat b}_{l+1})
({\hat X}_+(l))^{2\lambda_l}({\hat Y}_+(l))^{2\mu_{l}-1} \biggl]{\hat S}_0[l]
\nonumber\\
& &\ \ \ 
+\biggl[ \lambda_l ({\hat X}_+(l))^{2\lambda_{l}-1}({\hat Y}_+(l))^{2\mu_l}
{\hat a}_l^*{\hat a}_{l+1}^* \nonumber\\
& &\ \ \ \ \ \ +\mu_l{\hat a}_l^*{\hat b}_{l+1}
({\hat X}_+(l))^{2\lambda_l}({\hat Y}_+(l))^{2\mu_{l}-1} \biggl]{\hat S}_-[l]
\nonumber\\
& &\ \ \ 
+\biggl[ \lambda_l ({\hat X}_+(l))^{2\lambda_{l}-1}({\hat Y}_+(l))^{2\mu_l}
{\hat b}_l^*{\hat b}_{l+1}^* \nonumber\\
& &\ \ \ \ \ \ +\mu_l{\hat b}_l^*{\hat b}_{l+1}
({\hat X}_+(l))^{2\lambda_l}({\hat Y}_+(l))^{2\mu_{l}-1} \biggl]{\hat S}_+[l] 
\ , \nonumber\\
& &
[\ {\rm the\ other\ combinatiuons} \ ] = 0 \ . 
\end{eqnarray}
The relations (\ref{38a}), (\ref{38b}) and (\ref{39}) lead to the fact that 
the state (\ref{35}) is the eigenstate of ${\hat {\mib S}}[l]^2$ 
with the eigenvalue $\sigma_l(\sigma_l+1)$ : 
\begin{equation}\label{40}
\sigma_l=\lambda_l+\mu_l \ . \qquad (l=2,3,\cdots, n-1)
\end{equation}
The relations (\ref{37}) and (\ref{40}) give us 
\begin{eqnarray}\label{41}
& &2\lambda_1=s_1+s_2-\sigma_2 \ , \qquad
2\mu_1=s_1-s_2+\sigma_2 \ , \nonumber\\
& &2\lambda_l=\sigma_l+s_{l+1}-\sigma_{l+1} \ , \qquad
2\mu_l=\sigma_l-s_{l+1}+\sigma_{l+1} \ , \quad 
(l=2,3,\cdots, n-2)\nonumber\\
& &2\lambda_{n-1}=\sigma_{n-1}+s_n-s\ , \qquad
2\mu_{n-1}=\sigma_{n-1}-s_n+s\ .
\end{eqnarray}
The set $(\lambda_1\mu_1\lambda_2\mu_2\cdots \lambda_{n-1}\mu_{n-1})$ is 
just $(\alpha)$. It should be stressed that ${\hat S}_+$, ${\hat X}_+(l)$ 
and ${\hat Y}_+(l)$ play a role of building blocks to construct the 
eigenstates in many $su(2)$-spin system.

\section{General case. II}

As was shown in \S 3, 
we could learn a method which enables us to obtain the eigenstates in 
many spin system under the coupling order (\ref{27}). 
However, there exist many other orders, for example, 
$\sum_{i=1}^{4}{\hat {\mib S}}(i)=({\hat {\mib S}}(1)+{\hat {\mib S}}(2))
+({\hat {\mib S}}(3)+{\hat {\mib S}}(4))$. 
Our next task is to obtain the eigenstates, as an example, under the order 
\begin{eqnarray}\label{42}
\sum_{i=1}^{n}{\hat {\mib S}}(i)&=&
\left[\cdots\left[[{\hat {\mib S}}(1)+{\hat {\mib S}}(2)]
+{\hat {\mib S}}(3)
\right]+\cdots +{\hat {\mib S}}(m)\right] \nonumber\\
& &+\left[\cdots\left[[{\hat {\mib S}}(m+1)+{\hat {\mib S}}(m+2)]
+{\hat {\mib S}}(m+3)\right]+\cdots +{\hat {\mib S}}(n)\right] \ . \ \ \ 
\end{eqnarray}
Then, it may be convenient to discuss the present problem by classifying whole 
$su(2)$-spins into two groups : 
$(1,m)$- and $(m+1,n)$-group. 
The former and the latter consist of the spins 
$({\hat {\mib S}}(1), \cdots , {\hat {\mib S}}(m))$ and 
$({\hat {\mib S}}(m+1), \cdots , {\hat {\mib S}}(n))$, respectively. 
For each group, we define the operators 
\begin{equation}\label{43}
{\hat {\mib S}}[l;m]=
\cases{
{\hat {\mib S}}[l] \ , & $(l=1,2,\cdots , m)$ \cr
{\hat {\mib S}}[l]-{\hat {\mib S}}[m] \ . & $(l=m+1, m+2, \cdots , n) $
}
\end{equation}
Here, ${\hat {\mib S}}[l]$ is given in the relation (\ref{30}). 
Clearly, we have 
\begin{equation}\label{44}
{\hat {\mib S}}[1;m]={\hat {\mib S}}(1) \ , \qquad
{\hat {\mib S}}[m+1;m]={\hat {\mib S}}(m+1) \ . 
\end{equation}
Further, ${\hat {\mib S}}[m;m]$ and ${\hat {\mib S}}[n;m]$ denote the total 
spins of two groups, respectively, and then, ${\hat {\mib S}}$ can be 
expressed as 
\begin{equation}\label{45}
{\hat {\mib S}}={\hat {\mib S}}[m;m]+{\hat {\mib S}}[n;m] \ . 
\end{equation}
For each group, we use the building blocks 
$({\hat X}_+(l),{\hat Y}_+(l))$ for $l=1,2,\cdots , m-1$ and 
$({\hat X}_+(l),{\hat Y}_+(l))$ for $l=m+1,m+2,\cdots , n-1$. 
In addition to the above building blocks, we introduce the following 
operators : 
\begin{subequations}\label{46}
\begin{eqnarray}
& &{\hat X}_+(m,n)={\hat a}_m^*{\hat b}_n^*-{\hat b}_m^*{\hat a}_n^* \ , 
\label{46a}\\
& &{\hat Y}_+(m,n)={\hat a}_m^*{\hat a}_n+{\hat b}_m^*{\hat b}_n \ . 
\label{46b}
\end{eqnarray}
\end{subequations}
The operators (\ref{46a}) and (\ref{46b}) play a role of connecting 
the two groups. 
With the aid of the above operators, we define the state 
\begin{eqnarray}\label{47}
& &
\ket{\lambda_1\mu_1\cdots \lambda_{m-1}\mu_{m-1}\lambda_{m+1}\mu_{m+1}\cdots 
\lambda_{n-1}\mu_{n-1}\lambda_{mn}\mu_{mn};ss_0} \nonumber\\
&=&({\hat S}_+)^{s+s_0}({\hat X}_+(1))^{2\lambda_1}({\hat Y}_+(1))^{2\mu_1}
\cdots ({\hat X}_+(m-1))^{2\lambda_{m-1}}({\hat Y}_+(m-1))^{2\mu_{m-1}}
\nonumber\\
& &\times ({\hat X}_+(m+1))^{2\lambda_{m+1}}({\hat Y}_+(m+1))^{2\mu_{m+1}}
\cdots
({\hat X}_+(n-1))^{2\lambda_{n-1}}({\hat Y}_+(n-1))^{2\mu_{n-1}}\nonumber\\
& &\times ({\hat X}_+(m,n))^{2\lambda_{mn}}({\hat Y}_+(m,n))^{2\mu_{mn}}
({\hat b}_n^*)^{2s} \ket{0} \ .
\end{eqnarray}
The part $({\hat X}_+(m,n))^{2\lambda_{mn}}({\hat Y}_+(m,n))^{2\mu_{mn}}
({\hat b}_n^*)^{2s}\ket{0}\ (=\ket{\lambda_{mn}\mu_{mn};s})$ 
in the state (\ref{47}) can be rewritten as 
\begin{eqnarray}\label{48}
\ket{\lambda_{mn}\mu_{mn};s}&=&
\frac{(2s)!(2\lambda_{mn})!}{[2(s-\mu_{mn})]!}\sum_{r=0}^{2\lambda_{mn}}
(-)^{r}\frac{1}{r!(2\lambda_{mn}-r)!} \nonumber\\
& &\times ({\hat a}_m^*)^{2\lambda_{mn}-r}({\hat b}_m^*)^{2\mu_{mn}+r}
({\hat a}_n^*)^{r}({\hat b}_n^*)^{2(\lambda_{mn}-\mu_{mn}+s)-r}
\ket{0} \ . 
\end{eqnarray}

Let us prove that the state (\ref{47}) satisfies our purpose. 
In the present system, there exist $2n$ operators which are mutually 
commuted : ${\hat S}(l)\ (l=1,2,\cdots ,n)$, 
${\hat {\mib S}}[l;m]^2\ (l=2,3,\cdots ,m,m+2,m+3,\cdots , n)$, 
${\hat {\mib S}}^2$ and ${\hat S}_0$. 
Connecting ${\hat S}(l)$, in addition to the relation (\ref{36a}), 
we have 
\begin{subequations}\label{49}
\begin{eqnarray}
& &[\ {\hat S}(m)\ , \ {\hat X}_+(m,n)\ ]=(1/2){\hat X}_+(m,n) \ , \quad
[\ {\hat S}(m)\ , \ {\hat Y}_+(m,n)\ ]=(1/2){\hat Y}_+(m,n) \ , \nonumber\\
& &[\ {\hat S}(n)\ , \ {\hat X}_+(m,n)\ ]=(1/2){\hat X}_+(m,n) \ , \quad
[\ {\hat S}(n)\ , \ {\hat Y}_+(m,n)\ ]=-(1/2){\hat Y}_+(m,n) \ .\nonumber\\
& &\label{49a}
\end{eqnarray}
Instead of the relation (\ref{36b}), we have 
\begin{equation}
{\hat S}(n)({\hat b}_n^*)^{2s}\ket{0}=s({\hat b}_n^*)^{2s}\ket{0} \ . 
\end{equation}
\end{subequations}
Then, the eigenvalues of ${\hat S}(l)$, which we denote as $s_l$, 
is obtained as 
\begin{eqnarray}\label{50}
& &s_1=\lambda_1+\mu_1 \ , \nonumber\\
& &s_l=(\lambda_{l-1}-\mu_{l-1})+(\lambda_l+\mu_l) \ , \qquad
(l=2,3,\cdots ,m-1) \nonumber\\
& &s_m=(\lambda_{m-1}-\mu_{m-1})+(\lambda_{mn}+\mu_{mn}) \ , \nonumber\\
& &s_{m+1}=\lambda_{m+1}+\mu_{m+1} \ , \nonumber\\
& &s_l=(\lambda_{l-1}-\mu_{l-1})+(\lambda_l+\mu_l) \ , \qquad
(l=m+2,m+3,\cdots ,n-1) \nonumber\\
& &s_n=(\lambda_{n-1}-\mu_{n-1})+(\lambda_{mn}-\mu_{mn})+s \ . 
\end{eqnarray}
In the case of ${\hat {\mib S}}[l;m]^2\ (l=2,3,\cdots, m-1,m+2,\cdots ,n-1)$, 
the relation obtained from the relation (\ref{39}) by replacing 
${\hat {\mib S}}[l]$ with ${\hat {\mib S}}[l;m]$ is available. 
Further, we have 
\begin{eqnarray}\label{51}
& &{\hat {\mib S}}[m;m]^2\ket{\lambda_{mn}\mu_{mn};s}
=(\lambda_{mn}+\mu_{mn})\ket{\lambda_{mn}\mu_{mn};s} \ , \nonumber\\
& &{\hat {\mib S}}[n;m]^2\ket{\lambda_{mn}\mu_{mn};s}
=(\lambda_{mn}-\mu_{mn}+s)\ket{\lambda_{mn}\mu_{mn};s} \ . 
\end{eqnarray}
With the use of the above relations, we can prove that the 
state (\ref{48}) is the eigenstate of ${\hat {\mib S}}[l;m]^2\ 
(l=2,3,\cdots, m, m+2, m+3, \cdots ,n)$ with the eigenvalue 
$\sigma_l(\sigma_l+1)$ : 
\begin{eqnarray}\label{52}
& &\sigma_l=\lambda_l+\mu_l \ , \qquad (l=2,\cdots,m-1,m+2,\cdots,n-1) 
\nonumber\\
& &\sigma_m=\lambda_{mn}+\mu_{mn} \ , \nonumber\\
& &\sigma_n=\lambda_{mn}-\mu_{mn}+s \ . 
\end{eqnarray}
The relations (\ref{50}) and (\ref{52}) give us 
\begin{eqnarray}\label{53}
& &2\lambda_1=s_1+s_2-\sigma_2 \ , \qquad 
2\mu_1=s_1-s_2+\sigma_2 \ , \nonumber\\
& &2\lambda_l=\sigma_l+s_{l+1}-\sigma_{l+1} \ , \qquad 
2\mu_l=\sigma_{l}-s_{l+1}+\sigma_{l+1} \ , \quad
(l=2,3,\cdots,m-1) \nonumber\\
& &2\lambda_{m+1}=s_{m+1}+s_{m+2}-\sigma_{m+2} \ , \qquad 
2\mu_{m+1}=s_{m+1}-s_{m+2}+\sigma_{m+2} \ , \nonumber\\
& &2\lambda_l=\sigma_l+s_{l+1}-\sigma_{l+1} \ , \qquad 
2\mu_l=\sigma_{l}-s_{l+1}+\sigma_{l+1} \ , \quad
(l=m+2,m+3,\cdots,n-1) \nonumber\\
& &2\lambda_{mn}=\sigma_{m}+\sigma_{n}-s \ , \qquad 
2\mu_{mn}=\sigma_{m}-\sigma_{n}+s \ . 
\end{eqnarray}
The set $(\lambda_1\mu_1\cdots \lambda_{m-1}\mu_{m-1}\lambda_{m+1}\mu_{m+1}
\cdots \lambda_{n-1}\mu_{n-1}\lambda_{mn}\mu_{mn})$ is just $(\alpha)$. 
We can understand that ${\hat S}_+$, ${\hat X}_+(l)$, ${\hat Y}_+(l)$, 
${\hat X}_+(m,n)$ and ${\hat Y}_+(m,n)$ play a role of building blocks to 
construct the eigenstates in many $su(2)$ spin system.

\section{Coupling rules for the two cases}

In this section, we contact with the coupling rules for the order (\ref{27}) 
and (\ref{42}). We discuss the order (\ref{27}) in detail. 
Since ${\hat S}_{\pm,0}$ commutes with any of ${\hat X}_+(l)$ and 
${\hat Y}_+(l)$, the state (\ref{35}) can be rewritten in the form 
\begin{eqnarray}\label{54}
\ket{\lambda_1 \mu_1 \cdots \lambda_{n-1} \mu_{n-1};ss_0}&=&
({\hat X}_+(1))^{2\lambda_1}({\hat Y}_+(1))^{2\mu_1}\cdots
({\hat X}_+(n-1))^{2\lambda_{n-1}}\nonumber\\
& &\times ({\hat Y}_+(n-1))^{2\mu_{n-1}}
({\hat S}_+)^{s+s_0}({\hat b}_n^*)^{2s} \ket{0} \ . 
\end{eqnarray}
For the state (\ref{54}), the index number in each operator should not be 
negative, and then, we have 
\begin{subequations}\label{55}
\begin{eqnarray}
& &2s \geq 0 \ , \qquad s+s_0 \geq 0 \ , 
\label{55a}\\
& &2\lambda_l \geq 0 \ , \qquad 2\mu_l \geq 0 \ . \quad 
(l=1,2,\cdots,n-1) 
\label{55b}
\end{eqnarray}
\end{subequations}
Total number of $({\hat a}_n^* ,{\hat b}_n^*)$ in the part 
$({\hat S}_+)^{s+s_0}({\hat b}_n^*)^{2s}\ket{0}$ is equal to 
$2s-(s+s_0)$, and then, we can set up 
\begin{subequations}\label{56}
\begin{equation}
2s-(s+s_0) \geq 0 \ . 
\label{56a}
\end{equation}
Total number of $({\hat a}_n^* ,{\hat b}_n^*)$ in the part 
$({\hat Y}_+(n-1))^{2\mu_{n-1}}({\hat S}_+)^{s+s_0}({\hat b}_n^*)^{2s}
\ket{0}$ is equal to 
$2s-2\mu_{n-1}$, and we have 
\begin{equation}
2s-2\mu_{n-1} \geq 0 \ . 
\label{56b}
\end{equation}
Further, total number of $({\hat a}_l^* ,{\hat b}_l^*)\ 
(l=1,2,\cdots,n-2)$ in the part 
$({\hat Y}\_(l))^{2\mu_{l}}({\hat X}_+(l+1))^{2\lambda_{l+1}}
({\hat Y}_+(l+1))^{2\mu_{l+1}}\cdots ({\hat X}_+(n-1))^{2\lambda_{n-1}}
({\hat Y}_+(n-1))^{2\mu_{n-1}}({\hat S}_+)^{s+s_0}({\hat b}_n^*)^{2s}
\ket{0}$ is equal to 
$2\lambda_{l+1}+2\mu_{l+1}-2\mu_{l}$, and then, we get 
\begin{equation}
2\lambda_{l+1}+2\mu_{l+1}-2\mu_l  \geq 0 \ . \quad
(l=1,2,\cdots,n-2)
\label{56c}
\end{equation}
\end{subequations}
With the use of the relations (\ref{55}) and (\ref{56}), together with 
the relation (\ref{41}), we arrive at the following coupling rule : 
\begin{subequations}\label{57}
\begin{eqnarray}
& &s \geq 0 \ , \qquad 
|s_0| \leq s \ , 
\label{57a}\\
& &|s_1-s_2| \leq \sigma_2 \leq s_1+s_2 \ , \nonumber\\
& &|\sigma_l-s_{l+1}| \leq \sigma_{l+1} \leq \sigma_l+s_{l+1} \ , 
\quad (l=2,3,\cdots,n-2) \nonumber\\
& &|\sigma_{n-1}-s_{n}| \leq \sigma_{n} \leq \sigma_{n-1}+s_{n} \ . 
\label{57b}
\end{eqnarray}
\end{subequations}
Under the argument similar to the above, we obtain the coupling rule 
for the order (\ref{42}) in the form 
\begin{subequations}\label{58}
\begin{eqnarray}
& &s \geq 0 \ , \qquad 
|s_0| \leq s \ , 
\label{58a}\\
& &|s_1-s_2| \leq \sigma_2 \leq s_1+s_2 \ , \nonumber\\
& &|s_{m+1}-s_{m+2}| \leq \sigma_{m+2} \leq s_{m+1}+s_{m+2} \ , 
\nonumber\\
& &|\sigma_{l}-s_{l+1}| \leq \sigma_{l+1} \leq \sigma_{l}+s_{l+1} \ , 
\quad (l=2,3,\cdots,m-1,m+2,m+3,\cdots,n-1) \nonumber\\
& &|\sigma_m-\sigma_n|\leq s \leq \sigma_m+\sigma_n \ . 
\label{58b}
\end{eqnarray}
\end{subequations}
We derive the coupling rules (\ref{57}) and (\ref{58}) under the 
condition that the states (\ref{27}) and (\ref{42}) should not vanish 
and the result is completely the same as that in the 
conventional coupling rule.

\section{Application---the case of the $so(5)$-algebra}

In order to demonstrate the usefulness of the idea presented in this paper, 
we apply it to the construction of the orthogonal set for the 
$so(5)$-algebra. In this paper, we do not contact with physics related to 
this algebra. As a possible application of Ref. \citen{KPY00}, one of the 
present authors (M.Y.), Kuriyama and Kunihiro presented the Schwinger boson 
representation for the $su(4)$-algebra and its subalgebra.\cite{YKK00} 
One of them is the $so(5)$-algebra. 
This algebra consists of ten generators, which we denote as 
${\hat D}_{\pm,0}^*$, ${\hat D}_{\pm,0}$, ${\hat I}_{\pm,0}$ and 
${\hat M}_0$ : 
\begin{subequations}\label{59}
\begin{eqnarray}
& &{\hat D}_+^* = -({\hat a}_+^*{\hat b}_+ + {\hat a}_-^*{\hat b}_-) \ , 
\qquad
{\hat D}_-^* = {\hat \alpha}_+^*{\hat \beta}_+ 
+ {\hat \alpha}_-^*{\hat \beta}_- \ , \nonumber\\
& &{\hat D}_0^* = (1/2)({\hat a}_+^*{\hat \beta}_+ + {\hat a}_-^*{\hat \beta}_-
+{\hat \alpha}_+^*{\hat b}_+ + {\hat \alpha}_-^*{\hat b}_-) \ , \nonumber\\
& &{\hat D}_+ = -({\hat b}_+^*{\hat a}_+ + {\hat b}_-^*{\hat a}_-) \ , 
\qquad
{\hat D}_- = {\hat \beta}_+^*{\hat \alpha}_+ 
+ {\hat \beta}_-^*{\hat \alpha}_- \ , \nonumber\\
& &{\hat D}_0 = (1/2)({\hat \beta}_+^*{\hat a}_+ + {\hat \beta}_-^*{\hat a}_-
+{\hat b}_+^*{\hat \alpha}_+ + {\hat b}_-^*{\hat \alpha}_-) \ , 
\label{59a}\\
& &{\hat I}_+ = -{\hat a}_+^*{\hat \alpha}_+ - {\hat a}_-^*{\hat \alpha}_- 
+{\hat \beta}_+^*{\hat b}_+ + {\hat \beta}_-^*{\hat b}_- \ , \nonumber\\
& &{\hat I}_- = -{\hat \alpha}_+^*{\hat a}_+ - {\hat \alpha}_-^*{\hat a}_- 
+{\hat b}_+^*{\hat \beta}_+ + {\hat b}_-^*{\hat \beta}_- \ , \nonumber\\
& &{\hat I}_0 = (1/2)({\hat a}_+^*{\hat a}_+ - {\hat \alpha}_+^*{\hat \alpha}_+
+{\hat a}_-^*{\hat a}_- - {\hat \alpha}_-^*{\hat \alpha}_- 
\nonumber\\
& &\qquad\qquad\quad 
+{\hat \beta}_+^*{\hat \beta}_+ - {\hat b}_+^*{\hat b}_+
+{\hat \beta}_-^*{\hat \beta}_- - {\hat b}_-^*{\hat b}_-) \ , 
\label{59b}\\
& &{\hat M}_0 = (1/2)({\hat a}_+^*{\hat a}_+ + {\hat \alpha}_+^*{\hat \alpha}_+
+{\hat a}_-^*{\hat a}_- + {\hat \alpha}_-^*{\hat \alpha}_- 
\nonumber\\
& &\qquad\qquad\quad 
-{\hat \beta}_+^*{\hat \beta}_+ - {\hat b}_+^*{\hat b}_+
-{\hat \beta}_-^*{\hat \beta}_- - {\hat b}_-^*{\hat b}_-) \ . 
\label{59c}
\end{eqnarray}
\end{subequations}
Here, $({\hat a}_{\pm} , {\hat a}_{\pm}^*)$, 
$({\hat b}_{\pm} , {\hat b}_{\pm}^*)$, 
$({\hat \alpha}_{\pm} , {\hat \alpha}_{\pm}^*)$ and 
$({\hat \beta}_{\pm} , {\hat \beta}_{\pm}^*)$ denote boson operators, totally 
eight kinds. 
The expression (\ref{59a}) is slightly different from that shown in 
Ref. \citen{YKK00}. 
In this paper, we omit the physical meaning of these operators. 
In associating with the above, we can construct two sets of the operators. 
First is as follows : 
\begin{eqnarray}\label{60}
& &{\hat R}_+ = {\hat a}_+^*{\hat a}_- + {\hat b}_+^*{\hat b}_- 
+{\hat \alpha}_+^*{\hat \alpha}_- + {\hat \beta}_+^*{\hat \beta}_- \ , 
\nonumber\\
& &{\hat R}_- = {\hat a}_-^*{\hat a}_+ + {\hat b}_-^*{\hat b}_+ 
+{\hat \alpha}_-^*{\hat \alpha}_+ + {\hat \beta}_-^*{\hat \beta}_+ \ , 
\nonumber\\
& &{\hat R}_0 = (1/2)({\hat a}_+^*{\hat a}_+ - {\hat a}_-^*{\hat a}_- 
+{\hat b}_+^*{\hat b}_+ - {\hat b}_-^*{\hat b}_- 
\nonumber\\
& &\qquad\qquad\quad
+{\hat \alpha}_+^*{\hat \alpha}_+ - {\hat \alpha}_-^*{\hat \alpha}_- 
+{\hat \beta}_+^*{\hat \beta}_+ - {\hat \beta}_-^*{\hat \beta}_-) \ . 
\end{eqnarray}
The set $({\hat R}_{\pm,0})$ forms the $su(2)$-algebra and it satisfies 
\begin{equation}\label{61}
[\ {\hat R}_{\pm,0}\ , \ \hbox{\rm any\ of\ the\ }so(5)\hbox{-{\rm generators}}
\ ]=0 \ .
\end{equation}
Second is defined in the following form : 
\begin{subequations}\label{62}
\begin{eqnarray}
& &{\hat T}_{\pm,0}={\hat t}_{\pm,0}+{\hat \tau}_{\pm,0} \ , 
\label{62a}\\
& &{\hat t}_+={\hat a}_+^*{\hat b}_-^* - {\hat a}_-^*{\hat b}_+^* \ , 
\qquad
{\hat \tau}_+={\hat \alpha}_+^*{\hat \beta}_-^* 
- {\hat \alpha}_-^*{\hat \beta}_+^* \ , \nonumber\\
& &{\hat t}_-={\hat b}_-{\hat a}_+ - {\hat b}_+{\hat a}_- \ , 
\qquad
{\hat \tau}_-={\hat \beta}_-{\hat \alpha}_+
- {\hat \beta}_+{\hat \alpha}_- \ , \nonumber\\
& &{\hat t}_0=(1/2)({\hat a}_+^*{\hat a}_+ + {\hat b}_-^*{\hat b}_- 
+{\hat a}_-^*{\hat a}_- + {\hat b}_+^*{\hat b}_+)+1 \ , \nonumber\\
& &{\hat \tau}_0=(1/2)({\hat \alpha}_+^*{\hat \alpha}_+ 
+ {\hat \beta}_-^*{\hat \beta}_- 
+{\hat \alpha}_-^*{\hat \alpha}_- + {\hat \beta}_+^*{\hat \beta}_+)+1 \ .
\label{62b}
\end{eqnarray}
\end{subequations}
The sets $({\hat t}_{\pm,0})$ and $({\hat \tau}_{\pm,0})$ obey the 
$su(1,1)$-algebras, respectively, and then, the set $({\hat T}_{\pm,0})$ 
also obey the $su(1,1)$-algebra. 
The set $({\hat T}_{\pm,0})$ obeys 
\begin{subequations}\label{63}
\begin{eqnarray}
& &
[\ {\hat T}_{\pm,0}\ , \ \hbox{\rm any\ of\ the\ }so(5)\hbox{-{\rm generators}}
\ ] =0 \ , 
\label{63a}\\
& &[\ {\hat T}_{\pm,0}\ , \ \hbox{\rm any\ of}\ {\hat R}_{\pm,0} \ ] =0 \ . 
\label{63b}
\end{eqnarray}
\end{subequations}
The Casimir operator ${\hat \Gamma}_{so(5)}$ can be expressed as 
\begin{subequations}\label{64}
\begin{eqnarray}
{\hat \Gamma}_{so(5)}&=&
2\left[ {\hat D}_0^*{\hat D}_0 + (1/2)\left({\hat D}_+^*{\hat D}_+
+{\hat D}_-^*{\hat D}_-\right) 
+{\hat D}_0{\hat D}_0^*
+ (1/2)\left({\hat D}_+{\hat D}_+^*
+{\hat D}_-{\hat D}_-^*\right)\right] \nonumber\\
& &+\left[{\hat I}_0^2+(1/2)\left({\hat I}_-{\hat I}_+
+{\hat I}_+{\hat I}_-\right)\right] \ . 
\end{eqnarray}
The operator ${\hat \Gamma}_{so(5)}$ can be rewritten as 
\begin{equation}
{\hat \Gamma}_{so(5)}={\hat {\mib R}}^2+{\hat {\mib T}}^2 - 2 \ , 
\hspace{2.3cm}
\label{64b}
\end{equation}
\end{subequations}
\vspace{-0.5cm}
\begin{subequations}\label{65}
\begin{eqnarray}
& &{\hat {\mib R}}^2={\hat R}_0^2+(1/2)({\hat R}_-{\hat R}_+ 
+ {\hat R}_+{\hat R}_-) \ , 
\label{65a}\\
& &{\hat {\mib T}}^2={\hat T}_0^2-(1/2)({\hat T}_-{\hat T}_+ 
+ {\hat T}_+{\hat T}_-) \ . 
\label{65b}
\end{eqnarray}
\end{subequations}
The operators ${\hat {\mib R}}^2$ and ${\hat {\mib T}}^2$ denote the 
Casimir operators for the $su(2)$- and the $su(1,1)$-algebra, respectively. 
The above is the basic framework of the Schwinger boson representation 
for the $so(5)$-algebra presented in Ref. \citen{YKK00}. 
The Casimir operator (\ref{64b}) tells us that the orthogonal set 
for the $so(5)$-algebra should be the eigenstate of 
${\hat {\mib R}}^2$ and ${\hat {\mib T}}^2$.

The present system consists of eight kinds of bosons, and then, 
the orthogonal set is specified by eight quantum numbers. 
For example, they are chosen as $(R(R+1),\ R_0)$ for 
$({\hat {\mib R}}^2,\ {\hat R}_0)$, $(T(T-1),\ T_0)$ for 
$({\hat {\mib T}}^2,\ {\hat T}_0)$ and four quantum numbers additional to 
the above. 
First, we search the eigenstate for $({\hat {\mib R}}^2,\ {\hat R}_0)$. 
As is clear from the form (\ref{60}), the present 
is a system consisting of four $su(2)$-spins. 
We apply the coupling (\ref{42}) for the case $(m=2,\ n=4)$. 
Then, the boson operators ${\hat a}_+$, ${\hat a}_-$, ${\hat b}_+$, 
${\hat b}_-$, ${\hat \alpha}_+$, ${\hat \alpha}_-$, ${\hat \beta}_+$ and 
${\hat \beta}_-$ read the following way : 
\begin{eqnarray}\label{66}
& &{\hat a}_+ \rightarrow {\hat a}_1\ , \qquad 
{\hat a}_- \rightarrow {\hat b}_1 \ , \qquad
{\hat b}_+ \rightarrow {\hat a}_2\ , \qquad 
{\hat b}_- \rightarrow {\hat b}_2 \ , \nonumber\\
& &{\hat \alpha}_+ \rightarrow {\hat a}_3\ , \qquad 
{\hat \alpha}_- \rightarrow {\hat b}_3 \ , \qquad
{\hat \beta}_+ \rightarrow {\hat a}_4\ , \qquad 
{\hat \beta}_- \rightarrow {\hat b}_4 \ .
\end{eqnarray}
Further, ${\hat {\mib R}}$ reads ${\hat {\mib S}}$. 
Under the definitions (\ref{31}) and (\ref{46}), we have 
\begin{subequations}\label{67}
\begin{eqnarray}
& &{\hat X}_+(1)={\hat a}_+^*{\hat b}_-^* - {\hat a}_-^*{\hat b}_+^* 
={\hat t}_+ \ , \qquad
{\hat Y}_+(1)={\hat a}_+^*{\hat b}_+ + {\hat a}_-^*{\hat b}_- 
=-{\hat D}_+^* \ , \nonumber\\
& &{\hat X}_+(3)={\hat \alpha}_+^*{\hat \beta}_-^* 
- {\hat \alpha}_-^*{\hat \beta}_+^* ={\hat \tau}_+ \ , \qquad
{\hat Y}_+(3)={\hat \alpha}_+^*{\hat \beta}_+ 
+ {\hat \alpha}_-^*{\hat \beta}_- ={\hat D}_-^* \ , 
\label{67a}\\
& &{\hat X}_+(2,4)={\hat b}_+^*{\hat \beta}_+^* - 
{\hat b}_-^*{\hat \beta}_+^* \ , \qquad
{\hat Y}_+(2,4)={\hat b}_+^*{\hat \beta}_+ + {\hat b}_-^*{\hat \beta}_- \ . 
\label{67b}
\end{eqnarray}
\end{subequations}
Then, the form (\ref{47}) is reduced to 
\begin{subequations}\label{68}
\begin{eqnarray}
& &\ket{\lambda_1\mu_1\lambda_3\mu_3
\lambda_{24}\mu_{24};RR_0} \nonumber\\
&=&({\hat R}_+)^{R+R_0}({\hat t}_+)^{2\lambda_1}(-{\hat D}_+^*)^{2\mu_1}
({\hat \tau}_+)^{2\lambda_3}({\hat D}_-^*)^{2\mu_3} 
\nonumber\\
& &\times ({\hat b}_+^*{\hat \beta}_-^* 
- {\hat b}_-^*{\hat \beta}_+^*)^{2\lambda_{24}}
({\hat b}_+^*{\hat \beta}_+ 
+ {\hat b}_-^*{\hat \beta}_-)^{2\mu_{24}}
({\hat \beta}_-^*)^{2R}\ket{0} \nonumber\\
&=&({\hat R}_+)^{R+R_0}(-{\hat D}_+^*)^{2\mu_1}
({\hat D}_-^*)^{2\mu_3} 
({\hat t}_+)^{2\lambda_1}({\hat \tau}_+)^{2\lambda_3}
\ket{\lambda_{24}\mu_{24};R} \ .
\label{68a}
\end{eqnarray}
Here, ${\hat t}_+$ and ${\hat \tau}_+$ commute with ${\hat D}_{\pm}^*$, 
respectively, and $\ket{\lambda_{24}\mu_{24};R}$ is defined as 
\begin{eqnarray}
\ket{\lambda_{24}\mu_{24};R}&=&
({\hat b}_+^*{\hat \beta}_-^* 
- {\hat b}_-^*{\hat \beta}_+^*)^{2\lambda_{24}}
({\hat b}_+^*{\hat \beta}_+ 
+ {\hat b}_-^*{\hat \beta}_-)^{2\mu_{24}}
({\hat \beta}_-^*)^{2R}\ket{0} \nonumber\\
&=&
({\hat b}_+^*{\hat \beta}_-^* 
- {\hat b}_-^*{\hat \beta}_+^*)^{2\lambda_{24}}
({\hat b}_-^*)^{2\mu_{24}}
({\hat \beta}_-^*)^{2(R-\mu_{24})}\ket{0} \ . 
\label{68b}
\end{eqnarray}
\end{subequations}
Clearly, the state (\ref{68a}) is the eigenstate of $({\hat {\mib R}}^2, 
{\hat R}_0)$ with the eigenvalue $(R,R_0)$. 
But, it is not the eigenstate of $({\hat {\mib T}}^2 , {\hat T}_0)$. 
For obtaining the eigenstate, we, first, note the relations 
\begin{equation}\label{69}
{\hat t}_-\ket{\lambda_{24}\mu_{24};R}
={\hat \tau}_-\ket{\lambda_{24}\mu_{24};R}=0 \ , \qquad\qquad\quad
\end{equation}
\vspace{-0.8cm}
\begin{subequations}\label{70}
\begin{eqnarray}
& &{\hat t}_0\ket{\lambda_{24}\mu_{24};R}
=t\ket{\lambda_{24}\mu_{24};R} \ , \nonumber\\
& &{\hat \tau}_0\ket{\lambda_{24}\mu_{24};R}
=\tau\ket{\lambda_{24}\mu_{24};R} \ , 
\label{70a}\\
& &t=\lambda_{24}+\mu_{24}+1 \ , \qquad \tau=\lambda_{24}-\mu_{24}+R+1 \ , 
\end{eqnarray}
i.e., 
\begin{equation}
\lambda_{24}=(1/2)(t+\tau-R)-1\ , \qquad \mu_{24}=(1/2)(t-\tau+R) \ .
\label{70b}
\end{equation}
\end{subequations}
Then, the state (\ref{68a}) can be regarded as the eigenstate for 
$({\hat {\mib t}}^2, {\hat t}_0)$ and $({\hat {\mib \tau}}^2, {\hat \tau}_0)$ 
with the eigenvalues $(t(t-1), t_0)$ and $(\tau(\tau-1), \tau_0)$, 
respectively. 
Of course, $\lambda_1$ and $\lambda_3$ should be equal to 
\begin{equation}\label{71}
\lambda_1=(1/2)(t_0-t) \ , \qquad \lambda_3=(1/2)(\tau_0-\tau) \ . 
\end{equation}
Following a method presented in Ref. \citen{NPPTY04}, we can find the 
eigenstate of $({\hat {\mib T}}^2,{\hat T}_0)$, \break
$\ket{\mu_1\mu_3 t \tau;RR_0TT_0}$ in the form 
\begin{eqnarray}\label{72}
& &\ket{\mu_1\mu_3t\tau;RR_0TT_0} \nonumber\\
&=&({\hat R}_+)^{R+R_0}({\hat T}_+)^{T_0-T}(-{\hat D}_+^*)^{2\mu_1}
({\hat D}_-^*)^{2\mu_3} \nonumber\\
& &\times ({\hat O}_+)^{T-t-\tau}({\hat b}_+^*{\hat \beta}_-^*
-{\hat b}_-^*{\hat \beta}_+^*)^{t+\tau-R-2}
({\hat b}_-^*)^{R+t-\tau}({\hat \beta}_-^*)^{R-t+\tau}\ket{0} \ .
\end{eqnarray}
Here, ${\hat O}_+$ is defined as 
\begin{equation}\label{73}
{\hat O}_+={\hat t}_+({\hat t}_0 + t +\varepsilon)^{-1} 
- {\hat \tau}_+ ({\hat \tau}_0 +\tau + \varepsilon)^{-1} \ . 
\end{equation}
The parameter $\varepsilon$ plays a role for avoiding null denominator. 
For the state (\ref{72}), we have the following condition : 
\begin{eqnarray}\label{74}
& &\mu_1,\ \mu_3=0,\ 1/2,\ 1,\ 3/2,\cdots \ , \nonumber\\
& &t,\ \tau=1,\ 3/2,\ 2, \cdots \ , \nonumber\\
& &R=0,\ 1/2,\ 1,\ 3/2,\cdots \ , \qquad 
R_0=-R,\ -R+1,\ \cdots , \ R-1,\ R \ , \nonumber\\
& &T=2,\ 5/2,\ 3, \ 7/2,\cdots \ , \qquad 
T_0=T,\ T+1,\ T+2, \cdots \ , \nonumber\\
& &|t-\tau| \leq R \leq t+\tau-2 \ , \nonumber\\
& &T \geq t+\tau \ . 
\end{eqnarray}
The relation (\ref{74}) is derived under the condition that the state 
(\ref{72}) should not vanish. 
The state (\ref{72}) can be reexpressed in the following form : 
\begin{eqnarray}\label{75}
& &\ket{dd_0,\delta\delta_0,RR_0,TT_0}
=({\hat T}_+)^{T_0-T}({\hat R}_+)^{R+R_0}({\hat D}_+^*)^{d+d_0}
({\hat D}_-^*)^{\delta+\delta_0}\ket{d,\delta,R,T} \ , \nonumber\\
& &\ket{d,\delta,R,T}=({\hat O}_+)^{T-2-(d+\delta)}
({\hat b}_-^*)^{R+(d-\delta)}({\hat \beta}_-^*)^{R-(d-\delta)}
\nonumber\\
& &\qquad\qquad\qquad\qquad
\times ({\hat b}_+^*{\hat \beta}_-^* 
- {\hat b}_-^*{\hat \beta}_+^*)^{(d+\delta)-R}\ket{0}\ . 
\end{eqnarray}
The relation between the quantum numbers in the states (\ref{72}) and 
(\ref{73}) are given as 
\begin{equation}\label{76}
t=d+1\ , \quad \tau=\delta+1\ , \quad 2\mu_1=d+d_0 \ , \quad 
2\mu_3=\delta+\delta_0 \ .
\end{equation}
Therefore, we have the restriction for $(d, d_0)$ and 
$(\delta, \delta_0)$ in the form 
\begin{eqnarray}\label{77}
& &|d-\delta| \leq R \leq d+\delta \ , \qquad
T\geq (d+\delta)+2 \ , \nonumber\\
& &d_0=-d\ , \ -d+1\ , \cdots,\ d-1\ , d \ ,\qquad
\delta_0=-\delta\ , \ -\delta+1\ , \cdots,\ \delta-1\ , \ \delta \ . \qquad\ \ 
\end{eqnarray}
The sets $({\hat D}_+^*, {\hat D}_+, (1/2)({\hat M}_0+{\hat I}_0))$ and 
$({\hat D}_-^*, {\hat D}_-, (1/2)({\hat M}_0-{\hat I}_0))$ 
form the $su(2)$-algebras which are mutually independent, and of course, 
they are also independent of $({\hat T}_{\pm,0})$ and $({\hat R}_{\pm,0})$. 
Then, we have 
\begin{eqnarray}\label{78}
& &{\hat D}_+\ket{d,\delta,R,T}={\hat D}_-\ket{d,\delta,R,T}
={\hat R}_-\ket{d,\delta,R,T}={\hat T}_-\ket{d,\delta,R,T}=0 \ , \nonumber\\
& &(1/2)({\hat M}_0+{\hat I}_0)\ket{d,\delta,R,T}=-d\ket{d,\delta,R,T} \ , 
\nonumber\\
& &(1/2)({\hat M}_0-{\hat I}_0)\ket{d,\delta,R,T}=-\delta\ket{d,\delta,R,T} 
\ , \nonumber\\
& &{\hat R}_0\ket{d,\delta,R,T}=-R\ket{d,\delta,R,T} \ , \nonumber\\
& &{\hat T}_0\ket{d,\delta,R,T}=T\ket{d,\delta,R,T} \ .
\end{eqnarray}
The quantum numbers $(d,d_0)$ and $(\delta,\delta_0)$ denote the quantum 
numbers characterizing the sets 
$({\hat D}_\pm^*, {\hat D}_\pm, (1/2)({\hat M}_0\pm{\hat I}_0))$.

\section{Discussions and concluding remarks}

Finally, we give three comments. As is well known, 
the orthogonal set for the $so(5)$-algebra is specified by six quantum 
numbers. 
On the other hand, the present case is specified by eight quantum numbers. 
The relations (\ref{61}) and (\ref{63}) tell us that any matrix element 
for the $so(5)$-generators does not depend on the quantum numbers $R_0$ 
and $T_0$. 
Therefore, it may be enough to consider the case $(R_0=-R\ , T_0=T)$ 
for the $so(5)$-algebra. 
This means that the orthogonal set is specified by six quantum numbers. 
This is the first comment. 

The second is related to the set $({\hat I}_{\pm,0})$. 
The operators ${\hat I}_{\pm,0}$ form the $su(2)$-algebra and 
the set $({\hat D}_{\pm,0}^*)$ is tensor with 
rank$=1$. 
It may be clear that the state (\ref{70}) is not eigenstate of 
$({\hat {\mib I}}^2 , {\hat I}_0)$. 
It is interesting to construct the eigenstate of 
$({\hat {\mib I}}^2, {\hat I}_0)$. The outline was sketched 
in Ref.\citen{22}. 
Combining the idea given in Ref.\citen{22} with that by Elliott,\cite{YKK00}
the following set is obtained : 
\begin{subequations}\label{79}
\begin{eqnarray}
& &\ket{TT_0RR_0;\nu JII_0}
=\left({\hat T}_+\right)^{T_0-T}\left({\hat R}_+\right)^{R+R_0}
\left({\hat {\mib D}}^{*2}\right)^\nu
\nonumber\\
& &\qquad\qquad\qquad\times \sum_{K_0 S_0}\langle JK_0RS_0 \ket{II_0}
Z_{JK_0}({\hat {\mib D}}^*)\ket{TRS_0} \ , 
\label{79a}\\
& &\ket{TRS_0}=({\hat \beta}_-^*)^{R+S_0}({\hat b}_-^*)^{R-S_0}
({\hat b}_+^*{\hat \beta}_-^* 
- {\hat b}_-^*{\hat \beta}_+^*)^{T-R-2}\ket{0}\ . 
\label{79b}
\end{eqnarray}
\end{subequations}
Here, ${\hat {\mib D}}^{*2}$ is a scalar for $({\hat I}_{\pm,0})$ defined 
by 
\begin{equation}\label{80}
{\hat {\mib D}}^{*2}={\hat D}_0^{*2}+(1/2)({\hat D}_+^*{\hat D}_-^*
+{\hat D}_-^*{\hat D}_+^*) \ . 
\end{equation}
The operator $Z_{JK_0}({\hat {\mib D}}^*)$ is related to the 
solid harmonics 
$Z_{JK_0}(r\theta\phi)=r^J Y_{JK_0}(\theta\phi)$ which is a 
homogeneous polynomial for 
$r_{\pm,0}$ with $r_{\pm}=x\pm iy=r\sin\theta \exp(\pm i\phi)$ and 
$r_0=z=r\cos\theta$. 
By replacing $r_{\pm,0}$ in the solid harmonics with 
${\hat D}_{\pm,0}^*$, we obtain $Z_{JK_0}({\hat {\mib D}}^*)$ 
which is of the rank $(J, K_0)$. The symbol 
$\langle JK_0RS_0\ket{II_0}$ denotes the Clebsch-Gordan coefficient. 
The state $\ket{TRS_0}$ satisfies the condition 
\begin{subequations}\label{81}
\begin{eqnarray}
& &{\hat D}_{\pm,0}\ket{TRS_0}=0 \ , 
\label{81a}\\
& &{\hat T}_-\ket{TRS_0}={\hat R}_-\ket{TRS_0}=0 \ , \nonumber\\
& &{\hat T}_0\ket{TRS_0}=T\ket{TRS_0} \ , \qquad
{\hat R}_0\ket{TRS_0}=-R\ket{TRS_0} \ , 
\label{81b}\\
& &{\hat {\mib I}}^2\ket{TRS_0}=R(R+1)\ket{TRS_0} \ , \qquad
{\hat I}_0\ket{TRS_0}=S_0\ket{TRS_0} \ , 
\label{81c}
\end{eqnarray}
\end{subequations}
\vspace{-0.8cm}
\begin{subequations}\label{82}
\begin{eqnarray}
& &{\hat M}_0\ket{TRS_0}=-m_0\ket{TRS_0} \ , 
\qquad\qquad\qquad\qquad\qquad\qquad\qquad
\label{82a}\\
& &m_0=T-2\ . 
\label{82b}
\end{eqnarray}
\end{subequations}
Clearly, $\ket{TT_0RR_0;\nu JII_0}$ is an eigenstate of 
${\hat {\mib T}}^2$, ${\hat T}_0$, ${\hat {\mib R}}^2$, ${\hat R}_0$, 
${\hat {\mib I}}^2$ and ${\hat I}_0$ with the eigenvalues $T(T-1)$, 
$T_0$, $R(R+1)$, $R_0$, $I(I+1)$ and $I_0$, respectively. 
Further, we note the relation 
\begin{subequations}\label{83}
\begin{eqnarray}
& &{\hat M}_0\ket{TT_0RR_0;\nu JII_0}=M_0\ket{TT_0RR_0;\nu JII_0}\ , 
\label{83a}\\
& &M_0=2\nu + J+m_0 \ . 
\label{83b}
\end{eqnarray}
\end{subequations}
It should be noted that the quantities $\nu$ and $J$ can run under the 
restriction independently. 
But, we have no relation to fix the values of $\nu$ and $J$, and then, the 
states (\ref{79}) are linearly independent, but may be, in general, 
not orthogonal. 
In this case, an appropriate method, for example, the Schmidt method, 
should be adopted. 
According to the charge independent pairing theory for the 
single-orbit (its angular momentum is $j$) shell model, the operator 
${\wtilde M}_0$ which corresponds to ${\hat M}_0$ is expressed as 
\begin{equation}\label{84}
{\wtilde M}_0=(1/2)({\wtilde N}-(2j+1)) \ . 
\end{equation}
Here, ${\wtilde N}$ denotes total fermion number operator. 
Therefore, for the system with ${\cal N}$ fermion number and $v$ seniority 
number, we have 
\begin{eqnarray}\label{85}
& &(1/2)({\cal N}-(2j+1))=2\nu + J-(T-2) \ , \nonumber\\
& &(1/2)(2j+1-v)=T-2 \ . 
\end{eqnarray}

Third is related to the state (\ref{72}) or (\ref{73}). 
Under a special device which is proper to the 
$so(5)$-algebra, we derive the state (\ref{72}) in Ref. \citen{YKK00} 
in different notations. 
However, the present one is derived in rather general framework. 
In this sense, it may be possible to apply the present idea to various cases.

\acknowledgement

The main part of this work was performed when the author M.Y. stayed 
at Coimbra in September of 2003 and 2004 and 
the author Y.T. in March and September of 2004. 
They express their sincere thanks to Professor J. da Provid\^encia, 
co-author of this paper, for his kind invitation.

\end{document}